\documentclass[prl,twocolumn,showpacs,preprintnumbers,amsmath,amssymb]{revtex4}
\usepackage{epsfig}
\usepackage{graphicx}
\usepackage{dcolumn}
\usepackage{bm}

\newcommand{\vf}{v_{\rm F}}

\newcommand{\ef}{\epsilon_{\rm F}}

\begin{document}

\title{Orthogonality catastrophe and Kondo effect in graphene.}
\author{Martina Hentschel}

\affiliation{Max Planck Institut f\"ur Physik komplexer Systeme.
N{\"o}thnitzer
  Str. 38. D-01187 Dresden. Germany.}

\author{Francisco Guinea}

\affiliation{Instituto de Ciencia de Materiales de Madrid. CSIC. Sor Juana
In\'es
  de la Cruz 3. E-28049 Madrid. Spain.}
\begin{abstract}
Anderson's orthogonality catastrophe in graphene, at energies close to the
Dirac
point, is analyzed. It is shown that, in clean systems, the orthogonality
catastrophe is suppressed, due to the vanishing density of states at the
Dirac point. In the presence of preexisting localized states at the Dirac
energy, the orthogonality catastrophe shows similar features to those found
in normal metals with a finite density of states at the Fermi level. The
implications for the Kondo effect induced by magnetic impurities, and for the
Fermi edge singularities in tunneling processes are also discussed.
\end{abstract}
\pacs{73.20.-r; 73.20.Hb; 73.23.-b; 73.43.-f, 72.15.Qm}

\maketitle
\section{Introduction.}
Graphene has attracted a great deal of attention recently, due to its novel
fundamental properties and potential
applications\cite{Netal04,Betal04,Netal05,Zetal05b,GN07}. It is by now 
well established that its electronic properties at low energies are well
described by the two dimensional Dirac equation. At half filling graphene
should be a semimetal, with a vanishing density of states. This fact implies
that many properties of a metal, which are parametrized by
the density of states at the Fermi level, are different in a clean graphene
sample. The description of the electronic bands in graphene based on the
Dirac equation 
also leads to localized states in samples with edges\cite{FWNK96,WS00} or
lattice defects\cite{VLSG05,Petal06}. These states change the density of
states near the Dirac energy, as they induce a peak at this energy. Hence,
the density of states of graphene at the Dirac energy can either vanish, in a
clean sample, or diverge, if localized states are induced.,

We study here Anderson's orthogonality catastrophe (AOC)\cite{A67} in clean
and dirty graphene. The AOC can be considered the simplest non trivial
feature in the response of a metal, and
it is dependent on the value of the density of states at the Fermi level. The
AOC directly leads to many singularities in experiments which probe the
dynamical response of a metal,  like the Fermi edge singularity in X-ray
absorption\cite{ND69,M93}, 
and singularities in the transport properties of quantum dots and metallic
grains\cite{UG91,BHGS00,AL04}.  In graphene, the interplay
between the AOC and Coulomb blockade may be relevant for the analysis of
transport experiments on small quantum dots\cite{BYBBM05,GN07}.

The Kondo effect induced by magnetic impurities in metals can be seen as a
direct consequence of Anderson's orthogonality
catastrophe\cite{AYH70,AY71}. The coupling 
between the impurity spin and the conduction electrons can be divided into a
transverse term, $J_\perp$, which leads to spin-flip processes, and a
longitudinal term, $J_\parallel$, which induces an AOC associated to the same
spin flips.  This AOC leads to a strong suppression of spin fluctuations,
although
the effects of $J_\perp$ prevail at the lowest temperatures. These two
competing processes can be defined, in a very transparent way, in the
dissipative two level system\cite{G85}, which is equivalent to the Kondo
Hamiltonian.
The Kondo temperature, $T_K$, can be seen as the scale at which
spin flip processes ultimately cut off the AOC.

The AOC is modified in disordered metals\cite{GBLA02} 
and ballistic mesoscopic systems, due to the changes in the
electronic wavefunctions\cite{HUB04,HUB05}. We will analyze the AOC in
graphene using the numerical methods explained in\cite{HUB04,HUB05} (see
also\cite{OT90}), and also a phaseshift analysis similar to that
in\cite{A67}.

We analyze first the phaseshifts induced by a local potential, first in clean
graphene, and then in graphene in the presence of preexisting localized
levels. The next section presents a numerical study of the full overlap
between the electronic ground state before and after the potential is turned
on, and an analysis of the scaling of this quantity with system size. The
last section discusses the main implications of our work for the Kondo effect
in graphene, and the Fermi edge singularities associated to tunneling
processes.

\section{Phaseshift analysis.}
The overlap $S$ between the Slater determinants which describe the electronic
wavefunction before and after a potential is switched on can be written
as\cite{A67}: 
\begin{equation}
S \le N^{\sum_l \frac{2 l + 1}{3 \pi^2} \sin^2 ( \delta_l )}
\label{overlap_0}
\end{equation}
where $N$ is the number of electrons, and $\delta_l$ is the phaseshift induced
by the potential in the scattered waves at the Fermi level with angular
momentum $l$. In a typical metal, a weak local potential of strength
$\epsilon_0 \ll \ef$ induces a phaseshift in the $s$ channel which can be
approximated by $\delta_0 \approx \epsilon_0 N (\ef ) \ll 1$, where $N (\ef
)$ is the density of states at the Fermi level.

This analysis can be extended in a straightforward way to
graphene, where the electronic wavefunctions can be approximated by the two
dimensional Dirac equation (see below):
\begin{equation}
{\cal H} \equiv \vf \left( \begin{array}{cc} 0 &\pm k_x +  i k_y \\ \mp k_x
+
    i k_y & 0 \end{array} \right)
\label{dirac}
\end{equation}
where the two signs correspond to the two inequivalent corners of the
Brillouin Zone of the honeycomb lattice.

We  use eq.(\ref{overlap_0}) in
order to describe the dependence of the overlap on the number of electrons,
by computing analytically the phaseshifts induced by different types of
potentials. In the following, we use energy and momentum units such that $\vf
= 1$. 
\subsection{Clean graphene.}
We analyze first the phaseshifts induced by a circular potential well in
clean graphene, and we describe the electronic wavefunctions using the
continuum Dirac equation, eq.(\ref{dirac}) . We expect that this
approximation will describe 
qualitatively the effects of a local perturbation in the graphene lattice.

We assume that the potential well can scatter electrons between the
$K$ and $K'$ valleys, as it is the case for sufficiently localized potentials
in graphene. 

Using cylindrical coordinates, the Hamiltonian in the
clean system can be written as:
\begin{widetext}
\begin{equation}
{\cal H} \equiv \left( \begin{array}{cccc} 0 &i e^{- i \phi} \partial_r +
    \frac{e^{- i \phi}}{r} \partial_\phi &0 &0 \\ i e^{i \phi}
\partial_r -
    \frac{e^{- i \phi}}{r} \partial_\phi &0 &0 &0 \\ 0 &0 &0
&- i e^{i \phi}
    \partial_r + \frac{e^{i \phi}}{r} \partial_\phi \\ 0 &0 &- i e^{-i
\phi}
    \partial_r - \frac{e^{-i \phi}}{r} \partial_\phi &0
\end{array} \right)
\label{hamil}
\end{equation}
\end{widetext}
where the two first entries correspond to the $K$ point, and the two last
ones to the $K'$ point.

We add a constant perturbation in the region $r \le R_0$:
\begin{equation}
V \equiv \left( \begin{array}{cccc} \epsilon_0 &0 &0 &\Delta \\ 0
&\epsilon_0
    &\Delta &0 \\ 0 &\Delta &\epsilon_0 &0 \\ \Delta &0
&0 &\epsilon_0
    \end{array} \right)
\label{potential}
\end{equation}
where $\epsilon_0$ is a constant energy shift, and $\Delta$ is a potential
which induces scattering between the two valleys, and it is compatible with
the symmetries of the honeycomb lattice\cite{MGV07}.  

We analyze the scattering of an incident $s$ wave with incoming energy $k$:
\begin{equation}
\Psi_{\rm inc} ( r , \phi ) \equiv \left( \begin{array}{c} J_0 ( k r ) \\ - i
    J_1 ( k r ) e^{i \phi} \\ 0 \\ 0 \end{array} \right)
\label{inc}
\end{equation}
where $J_0 ( x )$ and $J_1 ( x )$ are Bessel functions of the first
kind. They satisfy: $\lim_{x \rightarrow 0} J_0 ( x ) \approx 1$, and
$\lim_{x \rightarrow 0} J_1 ( x ) \approx x/2$. 

The reflected waves outside the well can be written as:
\begin{equation}
\Psi_{\rm ref} ( r , \phi ) \equiv R_1 \left( \begin{array}{c} Y_0 ( k r ) \\
    - i 
    Y_1 ( k r ) e^{i \phi} \\ 0 \\ 0 \end{array} \right) + R_2 \left(
    \begin{array}{c} 0 \\ 0 \\ i Y_1 ( k r ) e^{i \phi} \\ Y_0 ( k r )
\end{array} \right)
\label{ref}
\end{equation}
$Y_0 ( x )$ and $Y_1 ( x )$ are Bessel functions of the second
kind. They satisfy: $\lim_{x \rightarrow 0} Y_0 ( x ) \approx 2/\pi ( \log
(x/2) + \gamma )$, and
$\lim_{x \rightarrow 0} Y_1 ( x ) \approx - 2 / ( \pi x )$.
The first contribution on the right hand side of eq.(\ref{ref}) is a
reflected wave in the same valley, and the second term is a wave in the
opposite valley as the incident wave.

Inside the potential well, the spectrum has a gap for energies $\epsilon_0 -
\Delta \le \epsilon \le \epsilon_0 + \Delta$. Within this range of energies,
the wavefunction inside the well can be written as:
\begin{widetext}
\begin{equation}
\Psi_{\rm trans} ( r , \phi ) \equiv T_1 \left( \begin{array}{c}
    \frac{\sqrt{\Delta^2 - k'^2}}{\sqrt{2} \Delta} I_0 ( k' r ) \\
    + i  \frac{k'}{\sqrt{2} \Delta}
    I_1 ( k' r ) e^{i \phi} \\ 0 \\ \frac{1}{\sqrt{2}} I_0 ( k' r ) \end{array}
\right) + T_2 \left(
    \begin{array}{c} \frac{i k'}{\sqrt{2} \Delta} I_0 ( k' r ) \\
    \frac{\sqrt{\Delta^2 - k'^2}}{\sqrt{2} \Delta} I_1 ( k' r ) e^{i \phi} 
\\ i \frac{1}{\sqrt{2}} I_1 ( k' r ) e^{i \phi} 
 \\ 0 \end{array} \right)
\label{trans_1}
\end{equation}
$I_0 ( x )$ and $I_1 ( x )$ are modified Bessel functions of the first
kind. They satisfy: $\lim_{x \rightarrow 0} I_0 ( x ) \approx 1$, and
$\lim_{x \rightarrow 0} I_1 ( x ) \approx x/2$.
The value of $k'$ in eq.(\ref{trans_1}) is given by: $\epsilon =
  \sqrt{\Delta^2 - k'^2}$. As $k = \epsilon + \epsilon_0$, we 
have $k' = \sqrt{\Delta^2 - ( k - \epsilon_0 )^2 }$.

For $| \epsilon - \epsilon_0 | \ge \Delta$, we have:
\begin{equation}
\Psi_{\rm trans} ( r , \phi ) \equiv T_1 \left( \begin{array}{c}
    \frac{1}{\sqrt{2} \Delta} J_0 ( k' r ) \\
    - i \frac{\Delta}{\sqrt{2 ( \Delta^2 + k'^2 )}}
    J_1 ( k' r ) e^{i \phi} \\ 0 \\ \frac{\Delta}{\sqrt{2 ( \Delta^2 + k'^2 )}}
    J_0 ( k' r ) 
\end{array} \right) + T_2 \left(
    \begin{array}{c} \frac{i \Delta}{\sqrt{2 ( \Delta^2 + k'^2 )}} J_0 ( k' r )
\\
    \frac{1}{\sqrt{2}} J_1 ( k' r ) e^{i \phi} 
\\ \frac{\Delta}{\sqrt{2 ( \Delta^2 + k'^2 )}} J_1 ( k' r ) e^{i \phi} 
 \\ 0 \end{array} \right)
\label{trans_2}
\end{equation}
and $\epsilon = \sqrt{\Delta^2 + k'^2}$, and $k' = \sqrt{(k-\epsilon_0 )^2 -
\Delta^2}$.

The scattering phaseshifts are determined by the reflection coefficients
$R_1$ and $R_2$ defined in eq.(\ref{ref}). The boundary conditions at $r=R_0$
are simply the continuity of the spinors, which define a set of four
equations for the four variables $R_1 , R_2 , T_1$ and $T_2$. 

For $\Delta = 0$
we have $R_2 = T_2 = 0$ and $R_1 = \bar{R}$. As $\lim_{x \rightarrow \infty} J_0 (
  x ) \approx \sqrt{2/(\pi x)} \cos ( x - \pi/4 )$, and $\lim_{x \rightarrow
    \infty} Y_0 ( 
  x ) \approx \sqrt{2/(\pi x)} \sin ( x - \pi/4 )$,  the phaseshift $\delta$
  is $\tan ( \delta ) = \bar{R}$. We find:
\begin{equation}
\tan ( \delta ) = \bar{R} ( k R_0 )  = - \frac{J_1 [ ( k - \epsilon_0 ) R_0 ] J_0 ( k
  R_0 ) - J_0 [ ( k 
  - \epsilon_0 ) R_0 
  ] J_1 ( k R_0 )}{J_1 [ ( k - \epsilon_0 ) R_0 ] Y_0 ( k R_0 ) - J_0 [ ( k -
  \epsilon_0 ) R_0 
  ] Y_1 ( k R_0 )}
\label{d0}
\end{equation}
\end{widetext}
and:
\begin{equation}
\lim_{k R_0 \rightarrow \infty} \bar{R} ( k R_0 ) = \tan ( \epsilon_0 R_0 )
\label{limit}
\end{equation}

\begin{figure}[!t]
\begin{center}
\includegraphics[width=7cm,angle=-90]{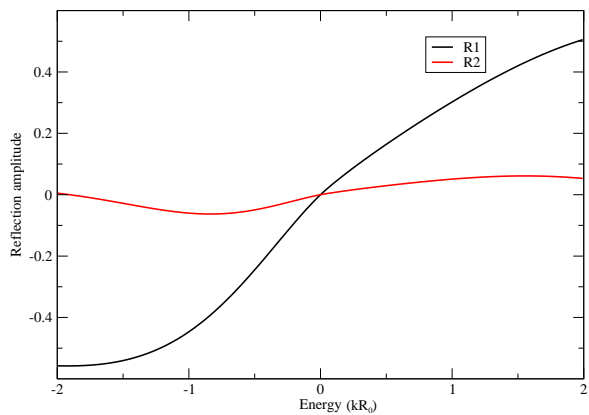}\\
\caption[fig]{\label{phaseshift} (Color online). Reflection coefficients of a
  circular well 
  with $\epsilon_0 R_0 = 0.5$ and $\Delta R_0 = 0.1$.}
\end{center}
\end{figure}

Results for $\epsilon_0 R_0 = 0.5$ and $\Delta R_0 = 0.1$ are shown in 
Fig.[\ref{phaseshift}]. In all cases, with or without (eq.(\ref{d0}))
intervalley scattering, 
the reflection coefficients vanish at the Dirac point, $k = 0$. This result
can be simply understood by noting that a finite reflection coefficient
implies a reflected wavefuntion with a component $Y_1 ( k r )$ which diverges
as $k \rightarrow 0$. The phaseshift vanishes linearly as $k \rightarrow 0$,
in agreement with general arguments based on the vanishing of the density of
states at the Dirac point. 

The vanishing of the phaseshift at the Dirac point  implies that the overlap
between the Slater determinants before and after the potential is switched on
does not scale like some power of the number electrons, and the AOC does not
take place at this energy.

\subsection{Phaseshift analysis in the presence of a localized state.}
\begin{figure}[!t]
\begin{center}
\includegraphics[width=6cm]{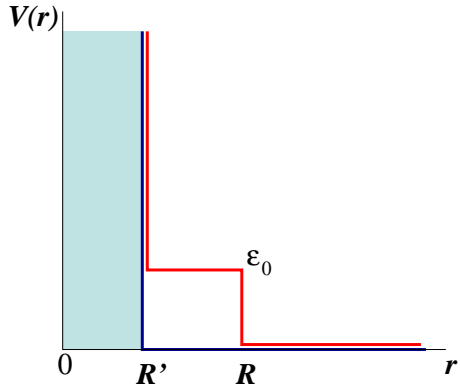}\\
\caption[fig]{\label{sketch_vacancy} (Color online). Sketch of the model with
  circular 
  symmetry used to study the AOC in the presence of localized levels. An
  infinite potential exists for $0 \le r \le R'$, mimicking a vacancy. The
  perturbation leading to the AOC is modeled as a constant potential,
  $\epsilon_0$, for $R' \le r \le R$ (see text for details).}
\end{center}
\end{figure}

 We will neglect here possible intervalley scattering terms. We study the
 phaseshifts induced by a weak potential near the edges of a circular void
which
 supports surface states. A sketch of the model is shown in
 Fig.[\ref{sketch_vacancy}]. We write the wavefunction as:
\begin{equation}
\Psi ( {\bf \vec{r}} ) \equiv \left( \begin{array}{c} \psi_1 ( {\bf \vec{r}} )
\\
    \psi_2 ( {\bf \vec{r}} ) \end{array} \right)
\end{equation} 
The edge of a crack, or extended vacancy is modeled by
the boundary condition:
\begin{equation}
\psi_1 ( {\bf \vec{r}} ) = 0 , \, \, \, \, \,  {\bf \vec{r}} \in \Omega
\label{boundary_void}
\end{equation}
where $\Omega$ is the boundary of the void.

We analyze a circular void, of radius $R'$. The boundary condition,
eq.(\ref{boundary_void}),  allows for
solutions at zero energy of the type:
\begin{equation}
\Psi ( {\bf \vec{r}} ) \equiv \left( \begin{array}{c} 0  \\
    \frac{e^{\pm i n \theta}}{r^n} \end{array} \right)
\end{equation}
where the two signs correspond to the two inequivalent corners of the
Brillouin zone.

Eq.(\ref{boundary_void}) implies,
for s-wave scattering:
\begin{equation}
\alpha_0 J_0 ( k R' ) + \beta_0 Y_0 ( k R' ) = 0
\end{equation}
The phaseshift induced by the void, before the potential whose effect we want
to calculate is turned on, is:
\begin{eqnarray}
\delta_0 ( k ) &= &\arctan \left( \frac{\beta_0}{\alpha_0} \right) = -
\arctan \left(
    \frac{J_0 ( k R' )}{Y_0 ( k R' )} \right) \nonumber \\ &\xrightarrow{k
\rightarrow 0}
  &- \frac{\pi}{2} \frac{1} {\ln ( k R' )} 
\label{phaseshift_void_0}
\end{eqnarray}

Next, we model a weak impurity near the void as an isotropic perturbation of depth
$\epsilon_0$, defined in the region $R' \le | {\bf \vec{r}} | \le R$. 
Following eqs.~(\ref{inc},\ref{ref}) and neglecting intervalley scattering,
the wavefunction can be written as:
\begin{widetext}
\begin{equation}
\Psi (  {\bf \vec{r}} ) \equiv \left\{ \begin{array}{lr} \left(
    \begin{array}{c}  \alpha' J_0 [ ( k + 
    \epsilon_0 ) r ] + \beta' Y_0 [ ( k + \epsilon_0 ) r ] \\ \alpha' J_1 [ (
    k + 
    \epsilon_0 ) r ] e^{i \phi } + \beta' Y_1 [ ( k + \epsilon_0 ) r ]  e^{i
    \phi }  \end{array} \right) &R'\le r \le R \\ 
 \left( \begin{array}{c}  \alpha J_0 (  k  r
    ) + \beta Y_0 ( k  r ) \\ \alpha J_1  ( k  
    r ) e^{i \phi } + \beta Y_1 ( k  r )  e^{i
    \phi }  \end{array} \right) &R \le r \end{array} \right.
\end{equation}
with boundary conditions:
\begin{eqnarray}
\alpha' J_0 [ ( k + \epsilon_0 ) R' ] + \beta' Y_0 [ ( k + \epsilon_0 ) R' ]
&= &0 \nonumber \\
\alpha' J_0 [ ( k + \epsilon_0 ) R ] + \beta' Y_0 [ ( k + \epsilon_0 ) R ]
&=
&\alpha J_0 ( k  R ) + \beta Y_0 ( k  R ) \nonumber \\
\alpha' J_1 [ ( k + \epsilon_0 ) R ] + \beta' Y_1 [ ( k + \epsilon_0 ) R ]
&=
&\alpha J_1 ( k  R ) + \beta Y_1 ( k  R )
\end{eqnarray}
\end{widetext}
These equations allow us to obtain the phaseshift of the combined system,
void and circular impurity, as $\delta = \arctan ( \beta / \alpha )$. The
overlap between the Slater determinants before and after the impurity
potential is switched on, is
determined by the phase difference, $\delta - \delta_0$, where $\delta_0$ is
given in eq.(\ref{phaseshift_void_0}). 

\begin{figure}[!t]
\begin{center}
\includegraphics[width=8cm]{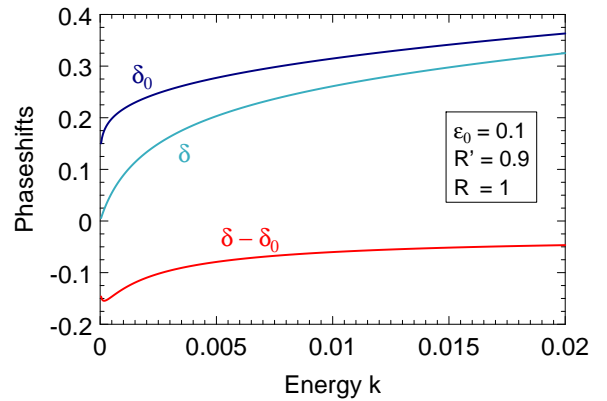}\\
\caption[fig]{\label{phaseshift_void_fig}
(Color online). Phaseshift $\delta_0$ induced by a void, 
the phaseshift $delta$ resulting from the additional switching on of a 
constant potential, and the resulting 
relative phaseshift $\delta - \delta_0$ 
induced by a circular impurity potential
surrounding a void (see text for details).}
\end{center}
\end{figure}

Results for the individual phaseshifts $\delta$ and $\delta_0$, 
as well as their difference are shown in
Fig.[\ref{phaseshift_void_fig}] for $\epsilon_0 = 0.1 , R' = 0.9$ and $R=1$. 
In this regime of energies much
lower than $\epsilon_0$, 
the phaseshift $\delta$ seems to approach $\delta_0$ from below, 
indicating that the repulsive character of the void is weakened by the 
additional constant potential.
For the small energies close to the Dirac point focused on here, the 
relative phaseshift, $\delta - \delta_0$, is always finite and 
seems to approach a constant.
This behavior differs strikingly from our findings for clean graphene where
the vanishing of the phaseshift at the Dirac point (cf. Fig.[\ref{phaseshift}])
indicates the suppression of AOC. 
In the presence of voids, the small dependence of the phaseshift 
induced by an additional external potential on
energy near the Dirac point implies that the overlap between Slater
determinants should scale with the number of electrons in a similar fashion
to that in a normal metal with a finite density of states. 
We shall see in the remainder of this paper that there are 
indeed considerable differences between clean graphene vs.~graphene with 
localized states, that are visible, e.g., in the behavior of the 
AOC overlap.

\section{Calculation of the overlap.}
\subsection{Clean graphene.}
\begin{figure}[!t]
\begin{center}
\includegraphics[width=8.5cm,angle=0]{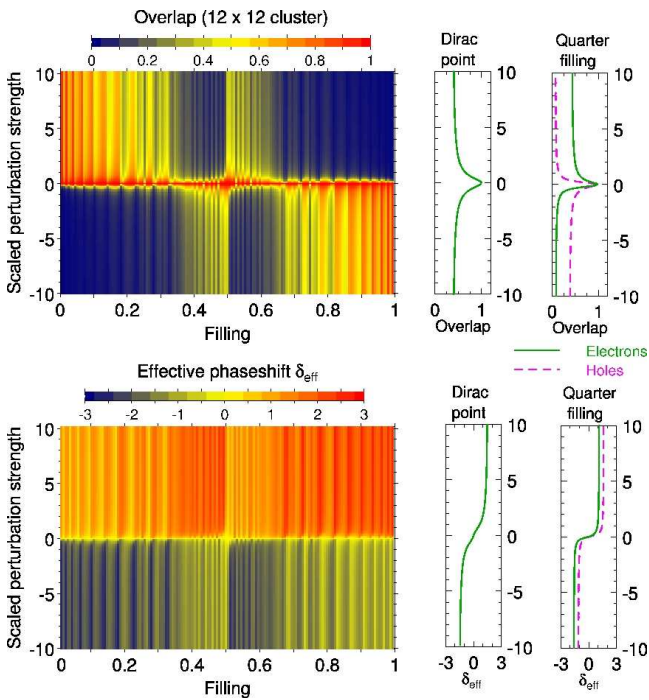}\\
\caption[fig]{\label{aoc_graph_fig}
(Color online). Overlap and effective phaseshift as function of filling and
potential strength (see
  text for details).}
\end{center}
\end{figure}
\begin{figure}[!t]
\begin{center}
\includegraphics[width=8cm,angle=0]{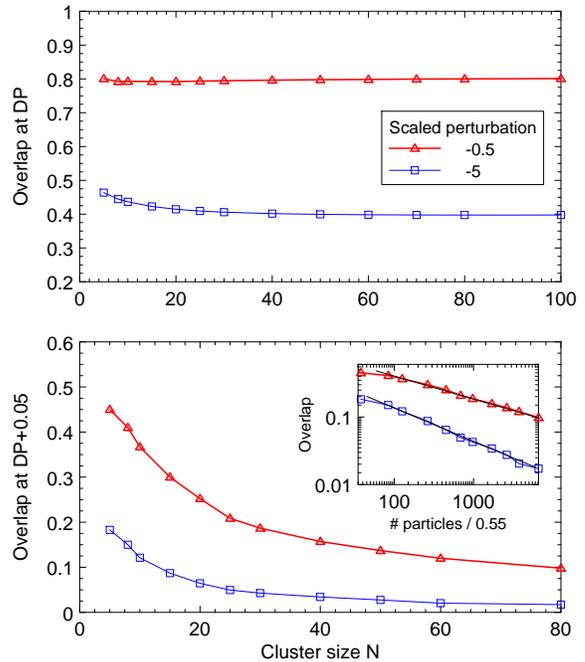}\\
\caption[fig]{\label{aoc_var_fig}
(Color online). Scaling of the overlap with cluster size $N$ 
at the Dirac point (DP, $\ef = 0$, 
corresponding
  to a filling of 0.5, top panel) and away from the Dirac point (at fixed filling
  0.55, corresponding to $\ef \sim 0.67$, lower panel). See 
  text for details.}
\end{center}
\end{figure}
The overlap between the unperturbed and perturbed Slater determinants for
clean graphene clusters of different sizes has been calculated using
the methods described in\cite{HUB04,HUB05,OT90}. 
The perturbation is a local potential at a given site, $\Delta = \epsilon_0$. 
Its strength is measured in terms of the scaled perturbation strength 
$\propto \Delta / d$ with $d$ being the mean level spacing 6/[(N (N+1)-2].
 Periodic
boundary conditions are used in systems with $N \times N$ unit cells, up to
$N=80$; the vertical stripes visible in Fig.[\ref{aoc_graph_fig}] are an
artefact of the periodic boundary conditions.
The results for the overlap for $N = 12$ and different potential
strengths (ranging from weak to strong for repulsive as well as 
attractive perturbations) are shown in
Fig.[\ref{aoc_graph_fig}]. 
An effective phaseshift can be defined by dividing
  the energy shift of the level closest to the Fermi energy by the average
  level spacing in that energy range. This phaseshift is also shown in
  Fig.[\ref{aoc_graph_fig}]. 

The dependence of the overlap with system size is different at the Dirac point
from that at other energies. This dependence is shown in
Fig.[\ref{aoc_var_fig}]. The overlap is almost independent of system size at
the Dirac point, cf.~the upper panel. 
This result is consistent with the phaseshift analysis,
which shows that the phaseshift vanishes at the Dirac point. Indeed, AOC is
suppressed at the Dirac point. Away from the Dirac point, the conventional
behavior
of the AOC overlap is recovered, see the lower panel of
Fig.[\ref{aoc_var_fig}].
To this end, AOC overlaps for fillings ranging from 0.54 to 0.56 were averaged
over.
Clearly, the AOC overlap is no longer suppressed and approaches zero in the
thermodynamic 
limit following the well-known power-law dependence on the number of particles 
($\propto [N (N+1)-2]$), cf.~inset of Fig.[\ref{aoc_var_fig}].

\subsection{Graphene with localized states.}
The method described in\cite{OT90} assumes that the wavefunctions
of all eigenstates of the unperturbed system have the same weight on the site
where the perturbation is turned on. This leads to a considerable
simplification of the calculation of the overlap between Slater
determinants. Generalization of this method 
generalized to 
to chaotic mesoscopic systems\cite{HUB04,HUB05} was done based on the
statistical properties of the chaotic wave functions.

In the presence of a defect which induces a localized state, like a vacancy,
the wavefunctions of the unperturbed state, where the localized state is
already present, do not possess translational symmetry. 
Therefore direct diagonalization and calculation
of the overlap determinants was used 
for the study of clusters of moderate sizes. 
\begin{figure}[!t]
\begin{center}
\includegraphics[width=7cm,angle=-90]{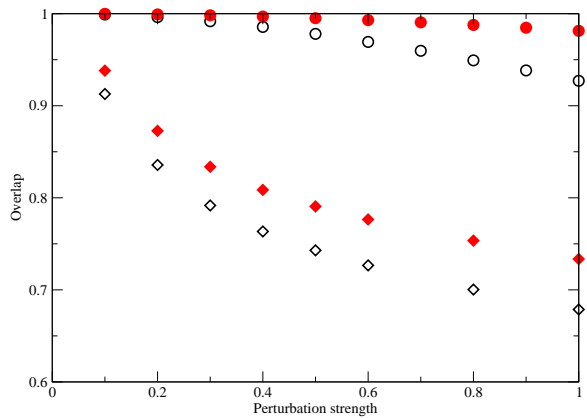}\\
\caption[fig]{\label{aoc_loc_fig}
(Color online). Dependence of the overlap on perturbation strength when the
  perturbation is 
  turned on near an existing vacancy (empty circles, black), and in
  clean graphene
  (filled circles, red). Calculations are done for $12 \times 12$ clusters.
  Circles
  correspond to one hole in the cluster (Dirac energy, $\ef = 0$), whereas 
  diamonds  
  characterize a cluster with five holes (corresponding to $\ef = - 0.5$, or a filling 
  of $\sim$ 0.47).}
\end{center}
\end{figure}
	
Results for the overlap for clusters with $12 \times 12$ unit
cells are shown 
in Fig.[\ref{aoc_loc_fig}]. At the Dirac point, the presence of a vacancy,
which induces a localized state, enhances significantly the dependence of the
overlap on the
strength of the potential. Away from the Dirac point, the difference in the
overlap with and without a vacancy is much less pronounced.
\begin{figure}[!t]
\begin{center}
\includegraphics[width=7cm,angle=-90]{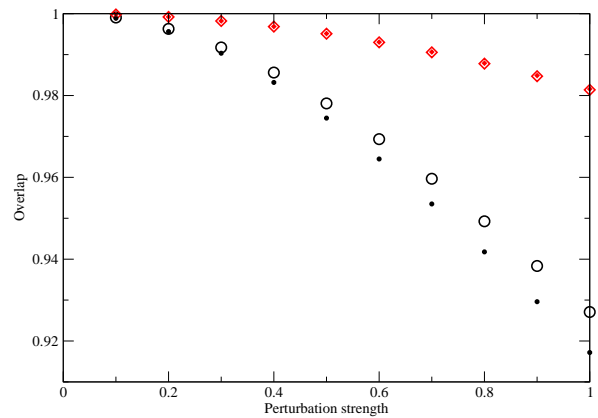}\\
\caption[fig]{\label{aoc_loc_2_fig}
(Color online). Dependence of the overlap on perturbation strength, at the
  Dirac energy, 
  when the perturbation is 
  turned on near an existing vacancy (circles, black), and in clean graphene
  (diamonds, red).  Large symbols
  correspond to a $12 \times 12$  cluster and small symbols correspond to a
  $15 \times 15$ cluster.}
\end{center}
\end{figure}
We show in Fig.[\ref{aoc_loc_2_fig}] the dependence of the overlap with
cluster size, at the Dirac energy. As anticipated in the discussion of 
Fig.[\ref{phaseshift_void_fig}],
the presence of a vacancy near the
potential which is turned on modifies significantly the results in comparison 
with a clean system.
In the latter, the dependence on size is negligible, in agreement with
the results shown in Fig.[\ref{aoc_var_fig}]. There is, on the other hand, a
substantial dependence on cluster size when a vacancy induces a localized
state at the Dirac energy.
\section{Conclusions.}
The results presented here show the existence of two regimes for  Anderson's
orthogonality 
catastrophe in 
graphene at low fillings, depending on whether there are localized states at
the Dirac energy or not. In the absence of localized states the AOC is
suppressed near the Dirac point, in agreement with the vanishing of the
density of states at this energy. When localized states are present, the AOC
is qualitatively similar to that found in metals with a finite density of
states. The latter behavior is a consequence of the fact that, when localized
states are sufficiently near the Fermi surface, they contribute to the non
adiabatic response of the electron gas. This situation is unique to graphene,
as, in most metallic systems, localized states appear at energies well below
the Fermi level.

The features discussed above imply that the Kondo effect in graphene also
depends on the strength of the scalar potential induced by the magnetic
impurity. If the potential induced on the graphene electrons is weak, as when
the magnetic impurity is at some distance of the graphene plane, we expect
the formation of a Kondo resonance to be suppressed, and the magnetic
impurity will give rise to a free magnetic moment. On the other hand, if the
magnetic impurity lies within the graphene plane, it will give rise to a
strong scalar potential, and possibly to localized states at the Dirac
energy. Then, the Kondo effect will not be suppressed, despite the low
density of states in graphene near the Dirac energy.

Similar effects can be expected for the Fermi edge singularities induced by
electrons tunneling into or out of graphene quantum dots. The strength of the
Fermi edge singularities depend on the existence of localized states in the
quantum dot. These states will be induced in graphene dots with sharp and
rough edges, where, in addition to Coulomb blockade, the AOC associated to
electron tunneling will further suppress the conductance at low
voltages\cite{UG91,BHGS00}. 
\section{Acknowledgments.}
F.~G.~acknowledges
funding from MEC (Spain) through grant FIS2005-05478-C02-01, the European
Union Contract 12881 (NEST), and CAM (Madrid) through 
program CITECNOMIK.
M.~H.~thanks the DFG (Germany) for funding within the Emmy-Noether program.

\bibliography{AOC_graphene}
\end{document}